\documentclass[11pt,a4paper]{article}
\usepackage{enumerate,amsthm,amsmath,graphicx,calc}
\usepackage{url}
\usepackage[lmargin=28mm,rmargin=28mm,tmargin=28mm,bmargin=28mm,footskip=10mm]{geometry}

\theoremstyle{plain}
\newtheorem{theorem}{Theorem}

\newcommand{\comment}[1]{~\newline\medskip\noindent\framebox{\parbox{\textwidth-2mm}{#1}}\medskip}

\renewcommand{\comment}[1]{}

\renewcommand{\baselinestretch}{1.2}
\begin{document}

\title{Faster Algorithms for Rigidity in the Plane}

\author{Sergey Bereg\thanks{
Department of Computer Science,
University of Texas at Dallas,
Box 830688,
Richardson, TX 75083,
USA. E-mail: {\tt besp@utdallas.edu}
}}

\maketitle




\begin{abstract}
In \cite{b-chclg-05}, a new construction called 
{\em red-black hierarchy} characterizing Laman graphs and an
algorithm for computing it were presented. 
For a Laman graph $G=(V,E)$ with $n$ vertices it runs in $O(n^2)$ time
assuming that a partition of $G+e,e\in E$ into two spanning trees is 
given. 
We show that a simple modification reduces the running time to
$O(n\log n)$. The total running time can be reduced 
$O(n\sqrt{n\log n})$ using 
the algorithm by Gabow and Westermann \cite{gw-ffgamsa-92} for
partitioning a graph into two forests. 
The existence of a red-black hierarchy is a necessary and
sufficient condition for a graph to be a Laman graph.
The algorithm for constructing a red-black hierarchy can be then
modified to recognize Laman graphs in the same time.
\end{abstract}

\section{Introduction}
We study minimally rigid graphs in the plane.
Several characterizations of these graphs are
known~\cite{w-rsa-04} including Laman counting 
(the graphs are also called {\em Laman graphs}) 
and tree partitions (partitions with two trees by Lov{\'a}sz and
Yemini and partitions with three trees by Crapo).
Tree partitions can be viewed as certificates for the property of a
graph being minimally rigid.
Recently, I found a new characterization of Laman graphs, a 
{\em red-black hierarchy} which is a hierarchical decomposition of a
graph. 

A red-black hierarchy of a graph $G=(V,E)$ can be constructed in
$O(n^2)$ time 
\cite{b-chclg-05}. The first step of the construction uses a partition
of $G+e,e\in E$ into two spanning trees. Using the algorithm by Gabow
and Westermann \cite{gw-ffgamsa-92} this step actually can be done in
$O(n\sqrt{n\log n})$ time only. The remaining part is a recursive
top-to-bottom construction of a red-black hierarchy. We show a slight
modification 
for the processing of a vertex of the hierarchy such that the running
time reduces to $O(n\log n)$ only.
If $k$ children are added to a hierarchy vertex then $k$ trees are
processed. The main idea is that a processing of $k-1$ trees
suffices. 
Surprisingly, this saves a significant time in computation.

The existence of a red-black hierarchy is a necessary and
sufficient condition for Laman graphs.
We show that the algorithm for constructing a red-black hierarchy can be 
modified to recognize Laman graphs in the same time.
This is similar to a recent algorithm developed by Daescu and
Kurdia~\cite{dk07,dk08}. However their algorithm is different since
it uses segment trees. 

{\em Henneberg construction}. 
Daescu and Kurdia~\cite{dk08} pointed out difficulties in using
red-black hierarchies for computing a Henneberg construction.
In a final version, we show that a hierarchical approach can still 
be applied here.

\section{Preliminaries}

Let $G=(V,E)$ be a graph and let $T=(V_T,E_T)$ be a rooted tree whose
set of leaves is in one-to-one correspondence with $V$. 
For a vertex $v$ of $V$, let $\alpha(v)$ be the corresponding vertex
of $V_T$.  
A {\em hierarchy} $H(G,T,\alpha,\beta)$ is defined as a graph with the
set of vertices $V_T$ and the set of edges $E_T\cup \beta(E)$ where
$\beta$ is a map $\beta:E\to V_T\times V_T$ that maps an edge
$e=(u,v)$ to a {\em cross edge} of $H$ such that 
the endpoints of $\beta(e)$ are corresponding ancestors of $\alpha(u)$ and
$\alpha(v)$. 

A hierarchy $H(G,T)$ is called a {\em red-black hierarchy} if it
satisfies the following conditions. 
\begin{itemize}
\item {\em Root Rule}.
The root has exactly two children.
\item {\em Leaf Rule}.
A vertex $v$ is the only child of its parent if and only if $v$ is a
leaf. 
\item {\em Cross-edge Rule}.
The endpoints of every non-tree edge have the same grandparent but
different parents.
\item {\em Tree Rule}.
For any vertex $v$, the cross edges incident to grandchildren of $v$
form a tree connecting all grandchildren of $v$.
\end{itemize}

\begin{figure}[htb]
\centering\includegraphics{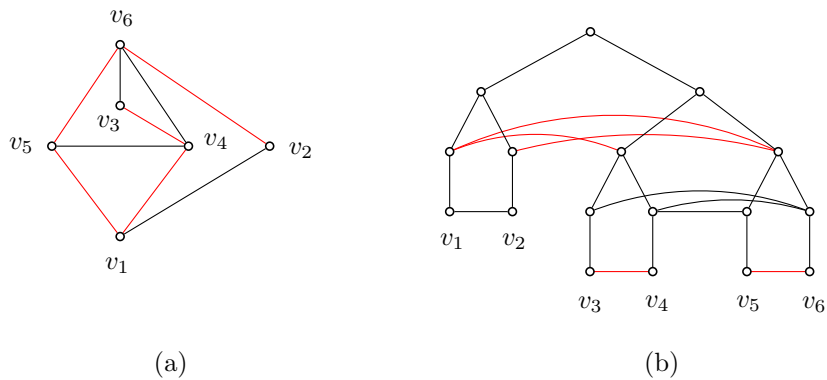}
\caption{(a) A Laman graph and (b) corresponding red-black
  hierarchy.}
\label{g1} 
\end{figure}

We denote by $|T|$ the {\em size} of a tree $T$, i.e. the number of its
vertices. 

The algorithm in \cite{b-chclg-05} first constructs two spanning trees
$T_R$ (red) and $T_B$ (black) of $G+e$ for an edge $e\in E$ and then
removes $e$ from $T_B$ making a forest $F_B$. The main step takes as
input a red tree $T_R$, black forest $F_B$ corresponding to a vertex
$v\in H$. Then 

1. Add $|F_B|$ children to $v$ so that each child $v_i$ corresponds to
a tree $T_i\in F_B$. Assign empty forest $F_i$ to each $v_i$.

2. Find red edges across the black forest (i.e. red edges whose
endpoints are in different trees of $F_B$).
Remove them from $T_R$ producing a red forest $F_R$.
Each red tree $T$ in $F_R$ connects vertices from the same black tree
$T_i$ and we add $T$ to $F_i$.
Proceed recursively with each child $v_i$, tree $T_i$ and forest $F_i$.

\begin{figure}[htb]
\centering\includegraphics{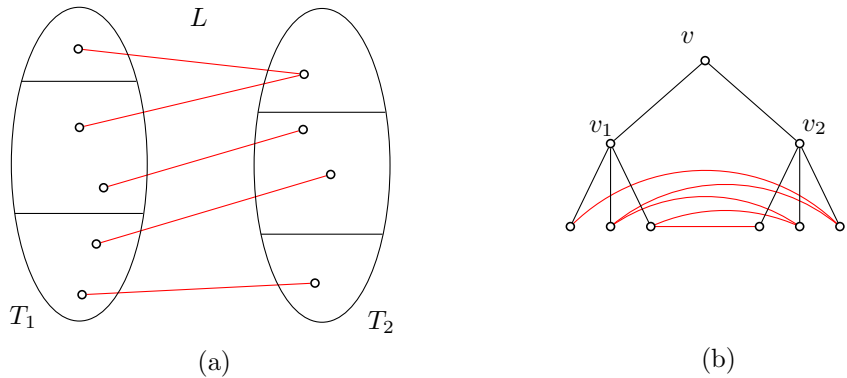}
\caption{(a) Trees $T_1,T_2$ of $F_B$. (b) The cross edges
  corresponding to $L$.}  
\label{g2} 
\end{figure}

\section{Faster Construction of a Red-black Hierarchy}

The main idea of improvement is to avoid unnecessary work in step
2. To find the list $L$ of red edges across the black forest, we
process 
vertices $v$ of all trees $T_i\in F_B$ except a maximum size tree
$T_j$. For each edge $e=(u,v)\in T_R$ incident to $v$, we add it to $L$
if $u$ is in a tree $T_{i'}$ different from $T_i$, i.e. $i\ne i'$.
To test an edge in $O(1)$ time, for each vertex, we store a
reference $tree(v)$ to a {\em tree header} which stores a name of the
tree and its size. 
The total number of tested edges for adding to $L$ is
$O(|T_R|-|T_j|+|L|)$. 
Indeed, the total number of vertices in the trees $T_i,i\ne j$ is
$|T_R|-|T_j|$. The number of edges with both endpoints in $T_i$ is
less than $|T_i|$. The total number of such edges is
$O(|T_R|-|T_j|)$. 
The time for computing $L$ is $O(|T_R|-|T_j|+|L|)$.

We show that the total time for computing lists $L$ over all recursive
calls is $O(n\log n)$. The total complexity of lists $L$ is 
$O(n)$. We charge the vertices of $T_i,i\ne j$ for the remaining time
$O(|T_R|-|T_j|)$. The size of each tree $T_i,i\ne j$ is at most
$|T_R|/k\le |T_R|/2$ where $k\ge 2$ is the number of trees in $F_B$.
The trees of $F_B$ were created by edge removals from a single black
tree in the previous level.
Thus, when a vertex is charged, its tree size has been
reduced by at least half. Then each vertex is charged 
$O(\log n)$ times in total. The running time follows.

{\em Edge removals}. We show that all edge removals can be done in
$O(n\log n)$ time. The removal of an edge $(u,v)$ from a tree $T$ 
results in two trees, say $T_u$ and $T_v$. 
We count the size of the smallest tree in $O(\min(|T_u|,|T_v|))$ 
time by parallel scanning the lists of vertices of $T_u$ and $T_v$. 
Without loss of generality $|T_u|\le |T_v|$ and, 
thus $|T_u|\le |T|/2$.  
We create a new tree header for $T_u$. 
Then update the sizes of $T_u$ and $T_v$ and tree references for the
vertices of $T_u$. The total time is $O(n\log n)$ since the tree
reference of every vertex changes $O(\log n)$ times.

\begin{theorem}
Let $G$ be a Laman graph with $n$ vertices. 
If a partition of $G$ into two forests is given then 
a red-black hierarchy for
$G$ can be constructed in $O(n\log n)$ time. 
\end{theorem}

\section{Testing Laman Graphs}
The above algorithm for constructing a red-black hierarchy assumes
that $G$ is a Laman graph. On the other hand the red-black hierarchy
is a certificate of Laman 
graphs and the algorithm can be modified to test whether $G$ is a
Laman graph. 

We use the algorithm by Gabow and Westermann \cite{gw-ffgamsa-92} for
partitioning a graph into two forests in 
$O(n\sqrt{n\log n})$ time.
The graph is not Laman if there is no such partition.
We run the above algorithm and test the red-black hierarchy rules on
the fly. 
The root rule is easy to check and we do it only for the root.

The current vertex $v$ has at least two children in
the hierarchy since $F_B$ has at least two trees. Thus the leaf rule
holds for the children of $v$. We check it for grandchildren of $v$. 
If a forest $F_i$ contains a single tree $T$ then $|T|$ must be one
(otherwise $G$ is not Laman and the algorithm stops).

The cross-edge rule and the tree rule for $v$ follows from the fact
that $T_R$ is a tree and previous checks.

\begin{theorem}
Let $G$ be a graph with $n$ vertices and $2n-3$ edges. 
If a partition of $G$ into two forests is given then 
it can be decided in
$O(n\log n)$ time whether $G$ is a Laman graph.
\end{theorem}

\end{document}